\documentstyle[aps,prl,multicol,epsf]{revtex}

\def \BE {\begin{equation}}
\def \EE {\end{equation}}
\def \BEA {\begin{eqnarray}}
\def \EEA {\end{eqnarray}}

\def\p {{\bf{ p}}}
\def\po {{\bf{ p_1}}}
\def\pt {{\bf{ p_2}}}
\def\pth {{\bf{ p_3}}}

\begin{document}
\title{Energy spectra of the ocean's internal wave field:
       theory and observations.}
 \author{Yuri V. Lvov$^1$, Kurt L. Polzin$^2$ and Esteban
G. Tabak$^3$,\\ \ \\
$^1$ Department of Mathematical Sciences, Rensselaer
Polytechnic Institute, Troy NY 12180\\ 
$^2$ Woods Hole Oceanographic
Institution, MS\#21, Woods Hole, MA 02543\\
$^3${Courant Institute of Mathematical Sciences,
        New York University, New York, NY 10012. }}
\maketitle
\pacs{91.10.V}
-\begin{abstract}
The high-frequency limit of the Garrett and Munk spectrum of internal
waves in the ocean and the observed deviations from it are shown to
form a pattern consistent with the predictions of wave turbulence
theory.  In particular, the high frequency limit of the Garrett and 
Munk spectrum constitutes an {\it exact} steady state solution of the 
corresponding kinetic equation.
\end{abstract}
\begin{multicols}{2}
{\bf Introduction. } 
Internal waves are an important piece of energy and momentum budgets
for the earth's atmosphere and ocean.  The drag associated with
internal wave breaking needs to be included in order to obtain accurate
simulations of the atmospheric Jet Stream \cite{FA03} and it has been
argued that the ocean's Meridional Overturning Circulation \cite{WF}
is forced by the diffusion of mass \cite{L00} associated with internal
wave breaking \cite{P97} rather than by the production of cold, dense
water by convection at high latitudes.  Both circulations represent
important pieces of the earth's climate system.

In a classical work \cite{GM72}, Garrett and Munk demonstrated how
observations from various sensor types could be synthesized into a
combined wavenumber-frequency spectrum, now called the
Garrett-and-Munk (GM) spectrum of internal waves. Consistent 
only with linear internal wave kinematics, the GM spectrum was
developed
as an  empirical curve fit to available data. Even though
deviations have been noted near boundaries \cite{WW79}, and at the
equator \cite{E85}, the last significant model revision
\cite{GM76,GM79} has surprisingly stood the test of time. However, a
review of open ocean data sets reveals subtle variability in spectral
power laws. We show in this letter that predictions based upon a
weakly nonlinear wave turbulence theory are consistent with both the
high frequency--high wave number limit of GM spectrum {\em and} the
observed variability.

In this letter, we will consider only the high frequency--high wave
number limit of GM; for brevity, we shall denote this henceforth as
GM$_{h}$.  The GM$_{h}$ is given by
\begin{equation}
  E(m,\omega) \simeq N m^{-2}\omega^{-2}.
\label{GM}
\end{equation}
Here $E$ is the spectral wave energy density, $N$ the buoyancy
frequency, $m$ the vertical wavenumber, and $\omega$ the frequency.
The total energy density of a wavefield is $E=\int E(m,\omega) ~ d m
d\omega$. 

The possibility that the internal wavefield might exhibit a universal
character represents an attractive theoretical target, and much effort
(as reviewed by \cite{M86}) was devoted to studying the issue of
nonlinearity in the context of resonant wave interactions.  That line
of work is based on a Lagrangian description of the flow, with two
main approximations: that fluid particles undergo small
displacements, and that nonlinear interactions take place on a
much longer time--scale than the underlying linear dynamics.  
An approximate kinetic equation describing the time evolution
of spectral wave energy was derived, and it was shown \cite{MM81} that
the GM$_{h}$ spectrum (\ref{GM}) was close to being a stationary solution.  

An alternative to the Lagrangian formulation, based on a Hamiltonian
description in isopycnal (density) coordinates, was recently proposed
\cite{LT}.  This approach does not invoke a small--displacement
assumption and yields a comparatively simple kinetic equation with an
exact steady power--law solution in the high frequency limit. That
steady state solution, [see (\ref{LTA}) below] is close to the
GM$_{h}$ spectrum (\ref{GM}), yet there is a noticeable difference.
Motivated by this difference, we tried to estimate the accuracy of the
GM$_h$ power laws and thus reviewed extant observations from the
literature. In the process of analyzing the data, we found that there
was subtle variability in the high wavenumber, high frequency
spectrum, forming a distinct pattern.

We then reexamined the kinetic equation of \cite{LT} and found its
full family of steady state solutions, of which the solution reported
in ~\cite{LT} is just one member. This family of solutions compares
well with the variability found in the observations. Moreover, the
GM$_{h}$ spectrum (\ref{GM}) is
a member of this family, thus describing the GM$_{h}$ spectrum simply as an
exact steady--state solution to the kinetic equation derived in
\cite{LT}.

Hence, in this article we present evidence for variability in the high
frequency-high wavenumber open-ocean internal wavefield, and find that
a wave turbulence approach predicts that both GM$_{h}$ itself and the
observed variability are stationary states of the kinetic equation.
The variability itself, and its likely roots in variable forcing,
Coriolis effects, underlying stratification and currents, as well as
the low frequency range of the energy spectrum, are fundamental
problems posing exciting challenges for future research.

{\bf Overview of observations: a family of spectra. } Below we
present a summary of historical oceanic internal wave
energy spectra. These observations are re-analyzed to
study whether the high-frequency, high-wavenumber spectra may form a
pattern. We review {\it seven} data sets available in the
literature. We shall present a detailed analysis of these data sets
elsewhere; here we just list them along with their high--frequency,
high--wave number
asymptotics.  Let us assume that, in this limit, 
the three dimensional wave action $n({\bf k},m)$ can be approximated by
horizontally isotropic power laws of the form
\begin{equation}
  n_{{\bf k},m}= n_0 \ |{\bf k}|^{-x} |m|^{-y} \, ,\label{PowerLawSpectrum}
\end{equation}
where ${\bf k}$ is the horizontal wave vector, 
$k=|{\bf k}|$ its modulus, $m$ vertical wavenumber and $n_0$ is a
constant.

Using the linear dispersion relation of internal waves, 
$\omega_{{\bf k},m}\propto |{\bf k}|/m$, this action spectrum can be
transformed from the wavenumber space $(k,m)$ to the vertical
wave-number-frequency space $(\omega,m)$ . Multiplication by the
frequency yields the corresponding energy spectrum,
$$ E(m,\omega)\propto \omega^{2-x} m^{2-x-y} ~~ .  $$ 
The total energy of the wave field is then
$$
E=\int \omega({\bf k},m) n({\bf k},m)\  d{\bf k} d m  =
\int  E(\omega,m) \ d\omega d m 
\label{TotalEnergy}. 
$$
Below we list extant data sets with concurrent vertical profile and
current meter observations and some major experiments utilizing moored
arrays, along with our best estimate
of their high-wave-number high frequency asymptotics (the order is
chronological):
\\ $\bullet $ The Mid-Ocean Dynamics Experiment (MODE), March-July 1973, 
Sargasso Sea ($26^{\circ} 0^{\prime}$ N, $69^{\circ} 40^{\prime}$ W): 
$m^{-2.25}\omega^{-1.6}$~\cite{L76};\\
$\bullet $ The Internal Wave Experiment (IWEX), 40 days observations
in November-December 1973, Sargasso Sea thermocline 
($27^{\circ} 44^{\prime}$ N, $69^{\circ} 51^{\prime}$ W):  
$k^{-2.4 \pm 0.4}\omega^{-1.75}$~ \cite{M78}; \\ 
$\bullet$ The Arctic Internal Wave Experiment (AIWEX),  March to May of
1985, Canada Basin thermocline, ($74^{\circ}$ N, $143-146^{\circ}$ W):
$m^{-2.15}\omega^{-1.2}$~\cite{Letal87,DandM91};\\
$\bullet $ The Frontal Air-Sea Interaction Experiment
(FASINEX), January to June of 1986, Sargasso Sea thermocline
($27^{\circ}$ N, $70^{\circ}$ W):  
$m^{-1.9~{\rm to}~-2.0}\omega^{-1.75}$~\cite{WandME,Eetal91};\\
$\bullet $ Patches Experiment (PATCHEX), 7.5 days during October of
1986, eastern Subtropical North Pacific, ($34^{\circ}$ N, $127^{\circ}$ W):  
$m^{-1.75}\omega^{-1.65~{\rm to}~-2.0}$~\cite{SandP91};\\
$\bullet $ The Surface Wave Process Program (SWAPP) experiment, 12
days during March, 1990, eastern Subtropical North Pacific thermocline, 
($35^{\circ}$ N, $127^{\circ}$ W):  $m^{-1.9}\omega^{-2.0}$~\cite{A92};\\
$\bullet $ North Atlantic Tracer Release Experiment (NATRE),
February-October 1992, eastern Subtropical North Atlantic thermocline,
($26^{\circ}$ N $29^{\circ}$ W): $m^{-2.75}\omega^{-0.6}$ (for $1 <
\omega < 6 $cpd)~\cite{P03}.\\
These deep ocean observations (Figure 1) exhibit a higher degree of
variability than one might anticipate for a universal spectrum.
Moreover, the deviations from the GM$_{h}$ spectral power laws form a
pattern: they seem to roughly fall upon a curve with negative slope in
the $(x,y)$ plane.  We show in the next section that the predictions of 
wave turbulence theory are consistent with this pattern.

{\bf A wave turbulence formulation for the internal wave field.} In
this section we assume that the internal wave field can be viewed as a
field of weakly interacting waves, thus falling into the class of
systems describable by wave turbulence.  Wave turbulence is a
universal statistical theory for the description of an ensemble of
weakly interacting particles, or waves. This theory has contributed to
our understanding of spectral energy transfer in complex
systems~\cite{Z92}, and has been used for describing surface water
waves since pioneering works by Hasselmann~\cite{H62}, Benney and
Newell \cite{BN} and Zakharov~\cite{Z68a,Z68b}.

The dynamics of oceanic internal waves can be most easily
described in isopycnal (i.e. density) coordinates, which allow for a
simple and intuitive Hamiltonian description~\cite{LT}. To describe
the wave field, we introduce two variables: a velocity potential
$\phi({\bf r},\rho)$, and an isopycnal straining $\Pi({\bf
r},\rho)$. The horizontal velocity is given by the {\it isopycnal} 
gradient $\nabla$ of the velocity potential,
${u({\bf r},\rho) =\nabla \phi({\bf r},\rho)}$.   
The straining
$\Pi =\rho/(\partial_z\rho)$ 
can also be interpreted as the fluid density in
isopycnal coordinates.

These two variables form a canonically conjugated Hamiltonian pair, so
that the primitive equations of motion (i.e. conservation of
horizontal momentum, hydrostatic balance, mass conservation and the
incompressibility constraint) can be written as a pair of canonical
Hamilton's equations,
\begin{eqnarray}
\partial_t \Pi=\delta {\cal H}/\delta \phi, \ \ \ \ \ \ \
\partial_t \phi=-\delta {\cal H}/\delta \Pi \, ,\nonumber\\
 {\cal H} = \frac{1}{2}\int \left( \Pi \, |\nabla \phi|^2 -
    \left|\int^{\rho} \frac{\Pi}{\rho_1} \, d\rho_1\right|^2 \right) 
d {\bf r} d \rho\,. \nonumber\\ \label{Hamiltonian}
\end{eqnarray}
The first term in the Hamiltonian clearly corresponds to the kinetic
energy of the flow, the second term can be shown to correspond
analogously to the potential energy.

Performing the Fourier transform, and introducing a complex field
variable $a_{\bf p}$ via
$$
\phi_\p=\frac{i N \sqrt{\omega_\p}}{\sqrt{2 g}k}
\left(a_\p-
a^*_{-\p}\right)\, ,
\Pi_\p =\frac{\sqrt{g} k}{\sqrt{2\omega_\p}N}\left(a_\p+a^*_{-\p}\right)\, 
, \nonumber
$$ where ${\bf p}=({\bf k},m)$ is the 3-d wave vector, $N$ is buoyancy
frequency, $g$ is gravity acceleration; the canonical pair of
equations of motion and the Hamiltonian (\ref{Hamiltonian}) read
\begin{eqnarray}
&&i\frac{\partial}{\partial t} a_{\p} = 
  \frac{\partial {\cal H}}{\partial a_{\p}^*},  \quad \hbox{where}\quad
{\cal H}=\int \omega_\p \, |a_{\p}|^2 \, d {\p} + \nonumber \\
&& \int
 V_{ {\po} {\pt} }^{\pth} \left( a_{\po}^* a_{\pt}^* 
a_{\pth} + a_{\po} a_{\pt}^* a_{\pth}^*\right)\,
\delta_{ {\po} -{\pt} - {\pth}} \, d{\po d \pt d \pth} +\nonumber \\
&&\int 
 \frac{1}{6}V_{ {\po} {\pt}}^ {\pth} \left( a_{\po}^* a_{\bf
p_2}^* a_{\pth}^* + a_{\po} a_{\pt} a_{\pth}\right) \,
\delta_{ {\po} + {\pt} +{\pth}}\, d  \po d \pt d \pth  ,\nonumber
\end{eqnarray}
with wave-wave interaction matrix elements given by ~\cite{LT}:
$V^{\p}_{\po\pt}= U^{\p}_{\po\pt}+U^{\po}_{\p\pt}+U^{\pt}_{\p\po}$
with
$$ 
U^{\p}_{\po\pt}= - \frac{N}{4\sqrt{2 g}}
\frac{{\bf k_2}\cdot{\bf k_3}}{k_2 k_3}
\sqrt{\frac{\omega_{\pt} \omega_\pth}{\omega_\po}} k_1 ~.
$$ 

These field equations are {\it equivalent} to the primitive equations
of motion for internal waves (up to the hydrostatic balance and
Boussinesq approximation); the work reviewed in ~\cite{M86} instead
resorted to a small displacement approximation to arrive at similar
equations. We will argue elsewhere that this extra assumption does not
provide an internally consistent description of interactions between
extremely scale separated waves.  For the purposes of this letter, it
suffices to note that the two kinetic equations are different and
yield different steady solutions.

We shall characterize the field of interacting internal waves by its
wave action
$\delta_{{\bf p} - { \bf p'}} n_{\bf p} = \langle a_{\bf p} a^*_{\bf
p'} \rangle.$

Under the assumption of weak nonlinear interaction, one derives a
closed equation for the evolution of the wave action, the kinetic
equation. Assuming horizontal isotropy, the kinetic equation can be
reduced further by averaging over all horizontal angles, obtaining
[with $p=(k,m)$ and $d p_1 d p_2 = d k_1 d m_1 d k_2 d m_2$]
\begin{eqnarray} 
 \frac{d n_{k,m}}{d t} 
= \frac{1}{k}\int 
\left(R^p_{p_1 p_2}-R^{p_1}_{p p_2}-R^{p_2}_{p_1 p}\right)
d p_1 d p_2  /\Delta^k_{k_1 k_2}, 
\nonumber \\
R^p_{p_1 p_2}= 
   \delta_{\omega_{p}-\omega_{p_1}-\omega_{p_2}} \,
f^p_{p_1 p_2} \, |V^p_{p_1 p_2}|^2 \, \delta_{m-m_1-m_2} k k_1 k_2 
\, , \nonumber\\
\label{KEaveraged}
\end{eqnarray}
where $f^p_{p_1 p_2} = n_{p_1}n_{p_2} - n_{p}(n_{p_1}+n_{p_2}) \ $ and   
$\Delta^k _{k_1 k_2} = \Big(
2 \left[ (k k_1)^2 +(k k_2)^2 +(k_1 k_2)^2 
\right]-k^4-k_1^4 -k_2^4\Big)^{1/2}/2$.

{\bf  A family of steady state power-law solutions to the kinetic
equation.}  In wave turbulence theory, three-wave kinetic equations
admit two classes of exact stationary solutions: thermodynamic
equilibrium and Kolmogorov flux solutions, with the latter
corresponding to a direct cascade of energy --or other conserved
quantities-- toward the higher modes.  The fact that the thermodynamic
equilibrium --or equipartition of energy-- $n_{\bf p} = 1/\omega_{\bf
p}$ is a stationary solution of (\ref{KEaveraged}) can be seen by inspection,
whereas in order to find Kolmogorov spectra one needs to be more
elaborate.  In \cite{LT} we used the Zakharov-Kuznetsov conformal
mapping \cite{Z68a,Z68b,Kuzia} to show analytically that the following
wave action spectrum constitutes an exact steady state solution of
(\ref{KEaveraged}) [note the difference with (\ref{GM})]:
\begin{eqnarray} n_{{\bf k},m}= n_0\  |{\bf k}|^{-7/2}
 |m|^{-1/2};\ 
E(m,\omega) \propto \omega^{-1.5}m^{-2}, \label{LTA}
\end{eqnarray} 
Remarkably though, this is not the only steady state solution of the
kinetic equation having nonzero spectral energy fluxes.  In fact, there
is a full family of such power law steady solutions.  To see this,
consider the kinetic equation (\ref{KEaveraged}), and substitute into
it the ansatz  (\ref{PowerLawSpectrum}).  Let us now denote the
resulting RHS of (\ref{KEaveraged}) by $I(k,m)$. For steady states,
$I(k,m)$ needs to vanish for all values of $k$ and $m$, for
appropriately chosen values of $(x,y)$. However, once $I$ vanishes for
one such wavenumber $(k,m)$, it does so for all, 
due to the fact that $I$ is a bi--homogeneous function of $k$ and $m$:
\begin{equation}\label{Rescaling}
I( \alpha  k ,\beta  m) = \alpha^{4 + 2 x} 
\beta^{1 + 2 y} I( k,  m).
\end{equation}
Hence we can fix $k$ and $m$, and seek zeros of $I$ as a function of
$x$ and $y$. The exact analytical solution (\ref{LTA}) cannot
correspond to an isolated zero of $I$, since 
$(\partial_x  I,\partial_y I)$ is
nonzero (it is proportional to the energy flux in the Kolmogorov
solution~\cite{Z92}). Hence, by the Implicit Function Theorem, there
must exist a curve of zeros of $I(x,y)$.

Since this family of steady state solutions is not all apparently
amenable to a closed form, we sought the zeros of $I$ by
numerical integration. This involves a certain amount of work. 
First, the delta-functions in (\ref{KEaveraged}) restrict
contributions to the resonant set. Consider, for example, the resonant
set
$${\bf k} = {\bf k_1} + {\bf k_2}, \  {m} = { m_1} + { m_2},\  
\omega_{k,m} = \omega_{k_1,m_1} + \omega_{k_2,m_2}.$$
\label{ResonantConditionM1}
\noindent 
Given $\bf k$, $\bf k_1$, $\bf k_2$ and $m$, one can find $m_1$ and
$m_2$ satisfying this resonant condition by solving
$$
{k}/{m}={k_1}/{|m_1|} +{k_2}/{|m-m_1|}.
$$
This equation reduces to a quadratic equation for $m_1$, and then one
can find $m_2$ from $m_2=m-m_1$.  After this reduction, one is left
with a two--dimensional integral, over $|k_1|$ and $|k_2|$. This
infinite domain is further restricted by the requirement that $|k_1|$,
$|k_2|$ and $|k|$ are such that they can correspond to the sides of a
triangle; this restricted (though still infinite) domain is called the
\emph{kinematic box} in the oceanographic literature.  The next
problem for the numerical integration is that the integrand diverges
(
typically in an integrable fashion) at the boundaries of the
domain. This is solved  by a suitable change of
variables. Finally, a second substitution renders the domain of
integration finite.

The resulting family of zeros is depicted in Figure 1.  Notice that
the curve passes through the exact solution~(\ref{LTA}). More
importantly, it also passes through the point $(4,0)$, corresponding
to the GM$_{h}$ spectrum (\ref{GM}). Hence this classical spectrum is for
the first time shown to correspond to an exact steady solution to a
kinetic equation based on first fluid principles.  

Finally, we note the integrals converge 
in the parameter regime occupied by the observations.  In regions of 
tightly spaced contour lines ($x < 1.7$ and $y< 0.7$, 
$x > 4.2$ and $y < -0.4$) (4) is nonintegrable.

The other points marked on the figure correspond to the observational
sets discussed above. Notice that, with the exception of NATRE, they
all lie very close to the zeros of $I$. 
Therefore the
predictions of wave turbulence are consistent with the
observed deviations from GM$_{h}$.

In fact, the NATRE point lies in an area of $(x,y)$
space where $z=I$ and $z=0$ are nearly tangential, thus making the
line of zeros effectively ``thicker'' (in other words, the collision
integral is not zero at the observed points, but it is very small,
possibly allowing other, typically smaller effects to take over.)

{\bf -- Conclusions} We have shown that the wave turbulence formalism
captures much of the variability apparent in the oceanic internal wave
field.  This includes the characterization of the spectral curve put
together by Garrett and Munk as an exact steady solution to a kinetic
equation for the evolution of the wave field, derived from first
principles.  In addition, the curve of steady solutions to this kinetic
equation is consistent with much of the observed variability
in the energy spectra.  We conjecture that the placement along this
curve of individual observations depends on the nature of the forcing
(for instance, by tides and atmospheric winds), the local degree of
stratification, vorticity and shear, and the variable magnitude of the
Coriolis parameter. This is the subject of ongoing research.

{\bf Acknowledgments} YL is supported by NSF CAREER grant DMS 0134955
and by ONR YIP grant N000140210528; KP is supported by NSF grant OCE
9906731; ET is supported by NSF grant DMS 0306729.
\begin{figure}
\epsfxsize=8.5cm
\epsffile{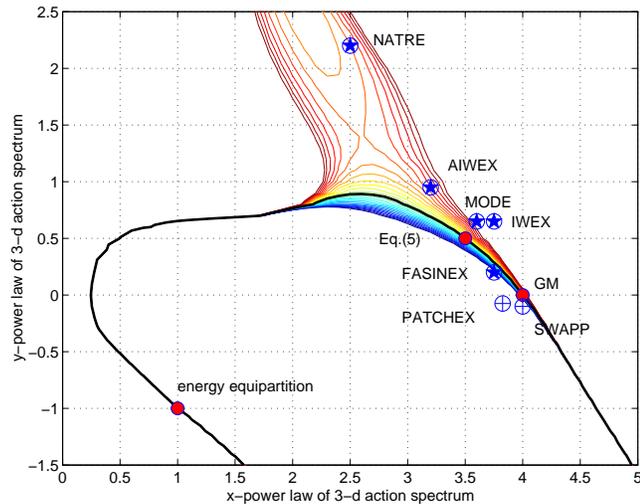}
\caption[]{\footnotesize\baselineskip11pt Ocean observations and
analytical zeroes of the kinetic equation (\ref{KEaveraged}) in the
$(x,y)$ plane, with the high--frequency action spectrum given by the
power law (\ref{PowerLawSpectrum}). Solid red dots represent the
thermodynamical equilibrium solution, the closed--form
zero~(\ref{LTA}) and the GM$_{h}$ spectrum~(\ref{GM}). Blue circles
represent different observational sets. The solid black curve marks
the numerically computed zeros of the kinetic equation. Contour lines
of the RHS of the equation (\ref{KEaveraged}) with high--frequency
action spectrum given by the power law (\ref{PowerLawSpectrum}),
$I(x,y)$, are also shown, with red curves correspond to positive
values, and blue to negative values.}
\end{figure}

\end{multicols}

\end{document}